\documentclass[english,twocolumn,showpacs,preprintnumbers,amsmath,amssymb]{revtex4}
\usepackage[T1]{fontenc}
\usepackage[latin1]{inputenc}
\usepackage{graphicx}

\makeatletter

\makeatletter

\usepackage{dcolumn}
\usepackage{bm}

\makeatother

\usepackage{babel}
\makeatother
\begin{document}

\title{Failure avalanches in fiber bundles for discrete load increase}

\author{Per C. Hemmer}

\email{per.hemmer@ntnu.no}

\author{Srutarshi Pradhan}

\email{pradhan.srutarshi@ntnu.no}

\affiliation{Department of Physics, Norwegian University of Science and Technology,
N--7491 Trondheim, Norway}

\begin{abstract}
The statistics of burst avalanche sizes $n$ during failure processes
in a fiber bundle follows a power law, $D(n)\sim n^{-\xi}$, for large
avalanches. The exponent $\xi$ depends upon how the avalanches are
provoked. While it is known that when the load on the bundle is increased
in a continuous manner, the exponent takes the value $\xi=5/2$, we
show that when the external load is increased in discrete and not
too small steps, the exponent value $\xi=3$ is relevant. Our analytic
treatment applies to bundles with a general probability distribution
of the breakdown thresholds for the individual fibers. The pre-asymptotic
size distribution of avalanches is also considered. 
\end{abstract}


\maketitle

\section{Introduction}

Fiber bundles with statistically distributed thresholds for breakdown
of individual fibers provide simple and interesting models of failure
processes in materials under stress. These models are much studied
since they can be analyzed to an extent that is not possible for more
complex materials (for reviews see \cite{Herrmann,Chakrabarti,Sornette,Sahimi,Pratip}).
We study here equal-load-sharing models, in which the load previously
carried by a failed fiber is shared equally by all the remaining intact
fibers in the bundle \cite{Peirce,Daniels,Smith,Phoenix}. The break-down
thresholds $x_{i}$ for the individual fibers are assumed to be independent
random variables with the same cumulative distribution function $P(x)$
and a corresponding probability density $p(x)$: \begin{equation}
\mbox{Prob}(x_{i}\leq x)=P(x)=\int_{0}^{x}p(y)\; dy.\end{equation}

We consider a bundle of $N$ fibers, clamped at both ends. At a force
$x$ per surviving fiber the total force $F(x)$ on the bundle is
$x$ times the number of intact fibers. The average, or macroscopic,
force is given by the expectation value of this, \begin{equation}
\langle F\rangle=N\; x\;[1-P(x)].\label{load}\end{equation}
 One may consider $x$ to represent the elongation of the bundle,
with the elasticity constant set equal to unity. In the generic case
$\langle F\rangle$ will have a single maximum $F_{c}$, a critical
load corresponding to the maximum load the bundle can sustain before
complete breakdown of the whole system. 

When a fiber ruptures somewhere, the stress on the intact fibers increases.
This may in turn trigger further fiber failures, which can produce
avalanches that either lead to a stable situation or to breakdown
of the whole bundle. One may study the burst distribution $D(n)$,
defined as the expected number of bursts of size $n$ when the bundle
is stretched until complete breakdown. When the load on the bundle
is increased continuously from zero, the generic result is a power
law \begin{equation}
\lim_{N\rightarrow\infty}\; N^{-1}\; D(n)\propto n^{-\xi},\end{equation}
 for large $n$, with $\xi=5/2$ \cite{Hemmer,Kloster}.

However, experiments may be performed in a different manner. In Sec.\ 2
we show for the uniform probability distribution of thresholds that
for experiments in which the load is increased in finite steps of
size $\delta$ rather than by infinitesimal amounts, the power-law
exponent is seen to increase to the value $\xi=3$. This have been
noticed in a special case previously: An argument by Pradhan \textit{et
al.} \cite{Pradhan5} for uniform threshold distribution suggested
this exponent value.

In Sec.\ 3 we show that the same asymptotic power law, \begin{equation}
D(n)\propto n^{-3}\label{-3}\end{equation}
 holds for a triangular threshold distribution and for a Weibull distribution.
The result (\ref{-3}) is demonstrated by simulations, and we provide
analytic derivations to back up the results. Our conclusions are not
limited to the distributions used in the simulations; we give an analytic
derivation valid for a {\em general} threshold distribution.

The avalanche distribution does not follow the asymptotic power law
(\ref{-3}) for small bursts. However, our analytic formulas do also
cover the distribution for bursts of smaller sizes $n$, down to a
minimum size.

In the concluding remarks we discuss briefly the results and their
range of applicability.

\section{Uniform threshold distribution}

For the uniform distribution of thresholds, \begin{equation}
P(x)=\left\{ \begin{array}{cl}
x & \mbox{ for }0\leq x\leq1\\
0 & \mbox{ for }x>1,\end{array}\right.\label{uniform}\end{equation}
 here in dimensionless units, the load curve (\ref{load}) is parabolic,
\begin{equation}
\langle F\rangle=N\; x\;(1-x),\label{loaduniform}\end{equation}
 so that the expected critical load equals \begin{equation}
F_{c}=N/4.\end{equation}

Fig.\ 1 shows the size distribution of the bursts obtained by simulation
on fiber bundles with the uniform threshold distribution (\ref{uniform}).
In one procedure the bursts are recorded under a continuous load increase,
in the other procedure the load is increased by discrete amounts $\delta$.
Clearly the exponent describing the large-size power laws are different,
close to $\xi=2.5$ and $\xi=3$, respectively.

\begin{center}\includegraphics[width=2.5in,height=2.5in]{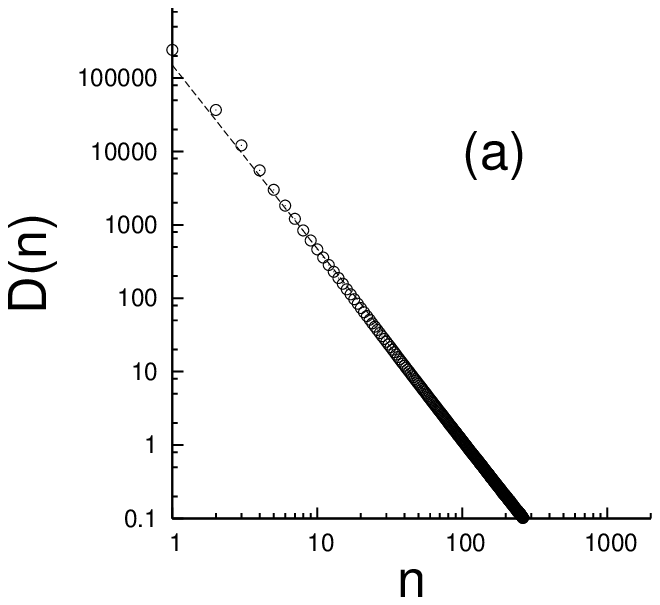}\par\end{center}

\begin{center}\includegraphics[width=2.5in,height=2.5in]{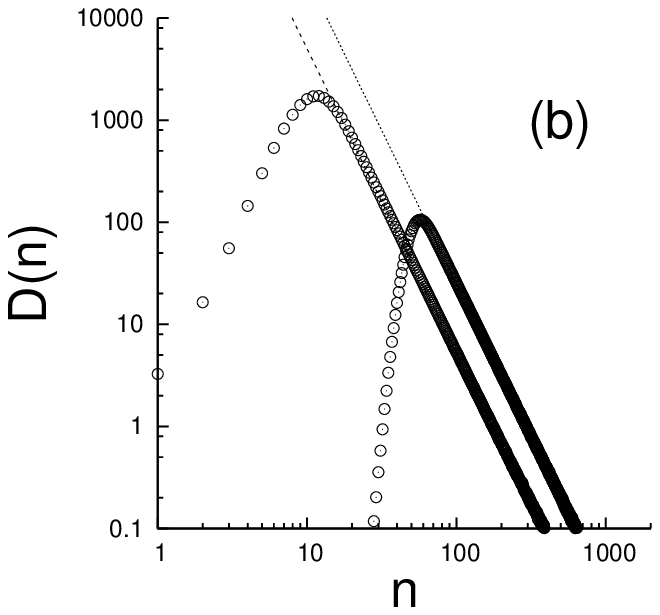}\par\end{center}

{\footnotesize FIG.\ 1. Avalanche size distribution for the uniform
threshold distribution (\ref{uniform}). The distribution marked (a)
is a record of all bursts when the load is increased continuously,
while the bursts recorded in (b) results from increasing the load
in steps of $\delta=10$ and $\delta=50$ (upper curve). The dotted
line in (a) represents a $n^{-5/2}$ behavior, and the dotted lines
in (b) show the theoretical asymptotics (\ref{Guniform}) for $\delta=10$
and $\delta=50$. The figure is based on $10000$ samples with $N=10^{6}$
fibers in the bundle.} \\

The basic reason for the difference in the power laws is that increasing
the external load in steps reduces the fluctuations in the force.
The derivation of the asymptotic size distribution $D(n)\propto n^{-5/2}$
of avalanches, corresponding to load increases by infinitesimal steps,
shows the importance of force fluctuations \cite{Hemmer}. An effective
reduction of the fluctuations requires that the size $\delta$ of
the load increase is large enough so that a considerable number of
fibers break in each step, i.e.\ $\delta\gg F_{c}/N$.

With a sufficiently large $\delta$ we may use the macroscopic load
equation (\ref{load}) to determine the number of fibers broken in
each step. The load values are $m\delta$, with $m$ taking the values
$m=0,1,2,\ldots,N/4\delta$ for the uniform threshold distribution.
By (\ref{loaduniform}) the threshold value corresponding to the load
$m\delta$ is \begin{equation}
x_{m}={\textstyle \frac{1}{2}}\left(1-\sqrt{1-4m\delta/N}\right).\end{equation}
 The expected number of fibers broken when the load is increased from
$m\delta$ to $(m+1)\delta$ is close to \begin{equation}
n=Ndx_{m}/dm=\delta/\sqrt{1-4m\delta/N}.\end{equation}
 Here the minimum number of $n$ is $\delta$, obtained in the first
load increase. The integral over all $m$ from $0$ to $N/4\delta$
yields a total number $N/2$ of broken fibers, as expected, since
the remaining one-half of the fibers burst in one final avalanche.

The number of avalanches of size between $n$ and $n+dn$, $D(n)\; dn$,
is given by the corresponding interval of the counting variable $m$:
\begin{equation}
D(n)\; dn=dm.\end{equation}

Since \begin{equation}
\frac{dn}{dm}=\frac{2\delta^{2}}{N}(1-4m\delta/N)^{-3/2}=\frac{2}{N\delta}\; n^{3},\end{equation}
 we obtain the following distribution of avalanche sizes: \begin{equation}
D(n)=\frac{dm}{dn}={\textstyle \frac{1}{2}}N\delta\; n^{-3},\hspace{1cm}(n\geq\delta).\label{Guniform}\end{equation}
 For consistency, one may estimate the total number of bursts by integrating
$D(n)$ from $n=\delta$ to $\infty$, with the result $N/4\delta$,
as expected.

Fig.\ 1 shows that the theoretical power law (\ref{Guniform}) fits
the simulation results perfectly for sufficiently large $n$. The
simulation records also a few bursts of magnitude less than $\delta$
because there is a nonzero probability to have bundles with considerably
fewer fibers than the average in a threshold interval. However, these
events will be of no importance for the asymptotic power law in the
size distribution. Moreover, for a more realistic threshold distribution
the situation is different (see below).

\section{General threshold distributions}

\begin{center}\includegraphics[width=2.8in,height=3in]{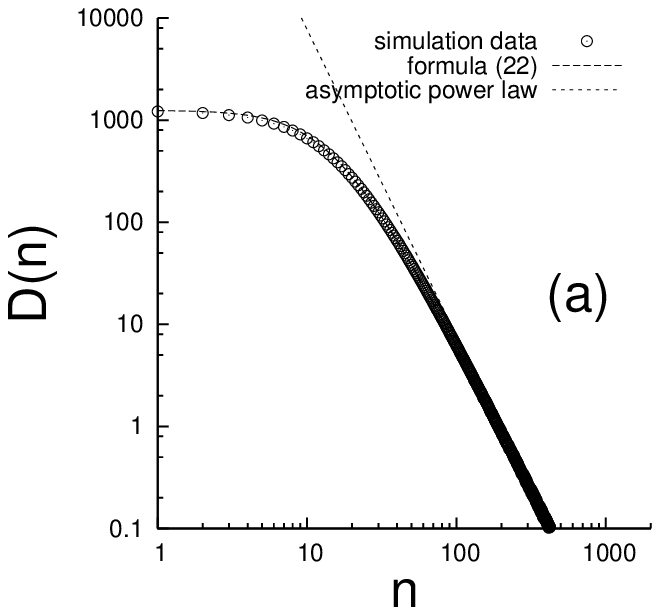}\par\end{center}

\begin{center}\includegraphics[width=2.8in,height=3in]{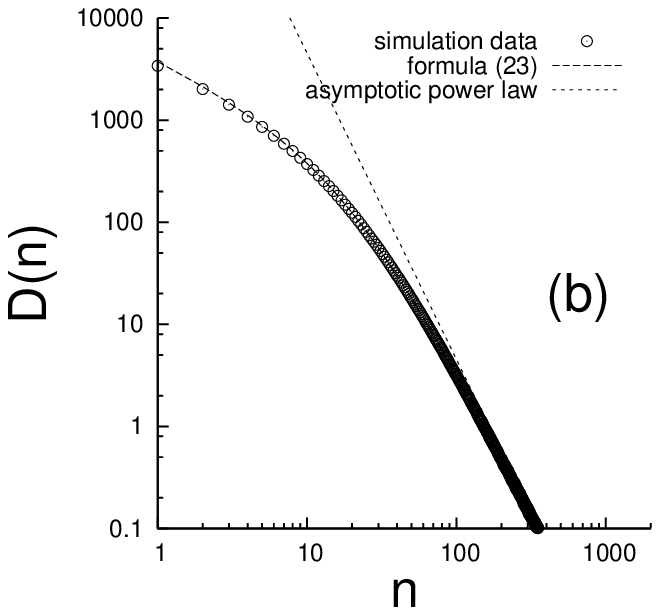}\par\end{center}

{\footnotesize FIG.\ 2 Avalanche size distribution for (a) the triangular
distribution (\ref{triangular}) and (b) for the Weibull distribution
(\ref{Weibull}). In both cases the load is increased in steps of
$\delta=20$. The figure is based on $10000$ samples of bundles with
$N=10^{6}$ fibers.}\\

In order to see whether the asymptotic law (\ref{-3}) is general,
we have performed simulations for two other threshold distributions,
the triangular threshold distribution \begin{equation}
p(x)=\left\{ \begin{array}{cl}
2x & \mbox{ for }0\leq x\leq1\\
0 & \mbox{ for }x>1,\end{array}\right.\label{triangular}\end{equation}
 and the Weibull distribution with index $5$, \begin{equation}
P(x)=1-e^{-x^{5}},\hspace{1.0cm}\mbox{ for all positive }x.\label{Weibull}\end{equation}

In Fig.\ 2 the exponent $\xi=3$ is clearly present in the triangular
distribution case, and less clearly in the Weibull distribution case.
Therefore we give now an analytic derivation for a general threshold
distribution, and apply the result to the two cases for which the
simulation results are shown in Fig.\ 2. The intention is to see
if the simulation results can be fitted, not merely with a power law
behavior for large avalanche sizes, but over an extended range of
$n$.

For a general threshold distribution $P(x)$ a load interval $\delta$
and a threshold interval are connected via the load equation (\ref{load}).
Since $d\langle F\rangle/dx=N[1-P(x)-xp(x)]$, an increase $\delta$
in the load corresponds to an interval \begin{equation}
\Delta x=\frac{\delta}{N[1-P(x)-xp(x)]}\label{dx}\end{equation}
 of fiber thresholds. The expected number of fibers broken by this
load increase is therefore \begin{equation}
n=N\; p(x)\;\Delta x=\frac{p(x)}{1-P(x)-xp(x)}\;\delta.\label{n}\end{equation}
 Note that this number diverges at the critical point, i.e.\ at the
maximum of the load curve, as expected.

We want to determine the number $D(n)\;\Delta n$ of bursts with magnitudes
in the interval $(n,n+\Delta n)$. The number $D(n)\;\Delta n$ we
seek is the number of load steps corresponding to the interval $\Delta n$.
Since each load step corresponds to the interval (\ref{dx}) of $x$,
the number of load steps corresponding to the interval $\Delta n$
equals \begin{equation}
D(n)\;\Delta n=\Delta n\;\frac{dx}{dn}\;\frac{N[1-P(x)-xp(x)]}{\delta}.\end{equation}

Using (\ref{n}) we have \begin{equation}
\frac{dn}{dx}=\frac{[1-P(x)]\; p'(x)+2p(x)^{2}}{[1-P(x)-xp(x)]^{2}}\;\delta.\end{equation}

Thus we obtain \begin{eqnarray}
D(n) & = & \frac{[1-P(x)-xp(x)]^{3}}{[1-P(x)]p'(x)+2p(x)^{2}}\;\frac{N}{\delta^{2}}\nonumber \\
 & = & \frac{p(x)^{3}}{[1-P(x)]p'(x)+2p(x)^{2}}\;\frac{N\delta}{n^{3}},\label{D(n)}\end{eqnarray}

\noindent using (\ref{n}). Here $x=x(n)$, determined by Eq.\ (\ref{n}).
For the uniform threshold distribution expression (\ref{D(n)}) coincides
with (\ref{Guniform}).

Near criticality the first fraction in the expression (\ref{D(n)})
for $D(n)$ becomes a constant. The asymptotic behavior for large
$n$ is therefore \begin{equation}
D(n)\simeq C\; n^{-3},\label{Dasymp}\end{equation}
 with a nonzero constant \begin{equation}
C=N\delta\;\frac{p(x_{c})^{2}}{2p(x_{c})+x_{c}p'(x_{c})}.\end{equation}
 We have used that at criticality $1-P(x_{c})=x_{c}p(x_{c})$. Thus
the asymptotic power law (\ref{-3}) is universal.

For the triangular threshold distribution considered in Fig.\ 2a
equations (\ref{n}) and (\ref{D(n)}) yield $n=2x\delta/(1-3x^{2})$
and $D=4N{\delta n}^{-3}x^{3}/(1+3x^{2})$. Elimination of $x$ gives
\begin{equation}
D(n)=\frac{2N\delta}{n^{3}}\;\;\frac{1}{6\delta/n+2\delta^{3}/n^{3}+(3+2\delta^{2}/n^{2})\sqrt{3+\delta^{2}/n^{2}}},\end{equation}
 with asymptotic behavior $D(n)\simeq(2N\delta/\sqrt{27})\; n^{-3}$,
in agreement with (\ref{Dasymp}).

For the Weibull distribution considered in Fig.\ 2b we obtain \begin{equation}
D(n)=N\delta\; n^{-3}\;\frac{25x^{9}\; e^{-x^{5}}}{4+5x^{5}},\hspace{.2cm}\mbox{ and }\hspace{.2cm}n=\frac{5\delta\; x^{4}}{1-5x^{5}}.\end{equation}
 This burst distribution must be given on parameter form, the elimination
of $x$ cannot be done explicitly. The critical point is at $x=5^{-1/5}$
and the asymptotics is given by (\ref{Dasymp}), with $C=N\delta(625e)^{-1/5}$.

These theoretical results are also exhibited in Fig.\ 2. In both
cases the agreement with the theoretical results is very satisfactory.
The $n^{-3}$ power law is seen, but the asymptotics sets in only
for very large avalanches, especially for the Weibull case. For small-sized
avalanches it is interesting to note the difference between the uniform
distribution and the more realistic Weibull distribution. In the latter
case the theoretical result (\ref{D(n)}) is reasonably accurate also
for $n<\delta$. The reason is that for the Weibull (and the triangular)
threshold distribution there are few very weak fibers. Thus a load
increase $\delta$ corresponds to a threshold interval that may, in
the average description, contain just a few failing fibers. For the
uniform distribution on the other hand, a corresponding threshold
interval with a number of fibers much less than $\delta$ can only
be caused by fluctuations with very low probabilities.

\section{Concluding remarks}

If we let the load increase $\delta$ shrink to zero, we must recover
the asymptotic $D(n)\propto n^{-5/2}$ power law valid for continuous
load increase. Thus, as function of $\delta$, there must be a crossover
from one behavior to the other. It is to be expected that for $\delta\ll1$
the $D(n)\propto n^{-5/2}$ asymptotics is seen, and when $\delta\gg1$
the $D(n)\propto n^{-3}$ asymptotics is seen.

We would like to emphasize that the step-wise load increase is a reasonable
as well as more practical loading method from the experimental point
of view. While performing fracture-failure experiments by applying
external load, one cannot ensure that a single fiber (the weakest
one) fails, whereas increasing the external load by equal steps is
an easier and more realistic procedure. 

In conclusion we have shown that the magnitude distribution $D(n)$
of avalanches generated by simulation for fiber bundles with step-wise
increase in the external load, is asymptotically a power law with
an exponent essentially equal to $-3$. We have analytically derived
this asymptotic power law $D(n)=Cn^{-3}$ for a general probability
distribution $p(x)$ of the individual fiber thresholds, as well as
the pre-asymptotic behavior of the avalanche distribution $D(n)$.

\vskip.2in

\textbf{\large Acknowledgment}{\large \par}

\vskip.1in

We would like to thank A. Hansen for useful comments. S. P. thanks
the Research Council of Norway (NFR) for financial support through
Grant No. 166720/V30 and 177958/V30.


\end{document}